# Joule Heating-Induced Particle Manipulation on a Microfluidic Chip

Golak Kunti,[a] Jayabrata Dhar,[b] Anandaroop Bhattacharya[a] and Suman Chakraborty[a]*

[a]Department of Mechanical Engineering, Indian Institute of Technology Kharagpur, Kharagpur, West Bengal - 721302, India

[b]Universite de Rennes 1, CNRS, Geosciences Rennes UMR6118, Rennes, France

*E-mail address of corresponding author: suman@mech.iitkgp.ernet.in

We develop an electrokinetic technique that continuously manipulates colloidal particles to concentrate into patterned particulate groups in an energy efficient way, by exclusive harnessing of the intrinsic Joule heating effects. Our technique exploits the alternating current electrothermal flow phenomenon which is generated due to the interaction between non-uniform electric and thermal fields. Highly non-uniform electric field generates sharp temperature gradients by generating spatially-varying Joule heat that varies along radial direction from a concentrated point hotspot. Sharp temperature gradients induce local variation in electric properties which, in turn, generate strong electrothermal vortex. The imposed fluid flow brings the colloidal particles at the centre of the hotspot and enables particle aggregation. Further, manoeuvering structures of the Joule heating spots, different patterns of particle clustering may be formed in a low power budget, thus, opening up a new realm of on-chip particle manipulation process without necessitating highly focused laser beam which is much complicated and demands higher power budget. This technique can find its use in Lab-on-a-chip devices to manipulate particle groups, including biological cells.

**INTRODUCTION**

Manipulation and assembly of colloidal particles including biological cells are essential in colloidal crystals, biological assays, bioengineered tissues, and engineered Laboratory-on-a-chip (LOC) devices.[1–6] Concentrating, sorting, patterning and transporting of microparticles possess several challenges in these devices/systems. For instance, sorting and concentrating of pathogenic cells at low concentrations is a difficult task which is essential for early disease detection.[7] To obviate such critical problems, numerous methods have been developed, namely dielectrophoresis,[8,9] magnetophoresis,[10] capillary electrophoresis,[11] evaporation-induced assembly,[12] optical tweezers,[13,14] Surface acoustic wave,[15] to name a few.

Because of several advantages such as, absence moving parts, easy integration with other components, target specific control etc dielectrophoresis (DEP) is one of popular technique for the manipulation of molecules and bio-particles in LOC devices.[16,17] In DEP the motion of a dielectric particle is generated through the interaction between applied non-uniform electric field and induced dipoles in the polarized particle. Magnetophoresis, on the other hand, manipulate magnetic particles using positive magnetophoresis. In this process, magnetic or magnetically tagged particles are transported along the magnetic field gradient.[18,19] Negative magnetophoresis is used to concentrate diamagnetic particles in both ferrofluids and paramagnetic solutions. Repulsion is adopted to manipulate diamagnetic particles. Magnetophoresis also posses several advantages, namely low operational cost, high throughput, less energy intensive, flexible for implementation, to name few.[20,21] Particle manipulation process using optical tweezers shows precise control. However, this process lacks high-throughput. Conventional optical tweezers aggregate the particles on the spot of a focused laser beam. Trapping volume of these tweezers is diffraction-limited.[22] On the other hand, nanometric optical tweezers deal with strong electromagnetic fields near the objects and recommend a subwavelength trapping volume.[23] Acoustic focusing method employs a ultrasonic field which generate acoustic radiation pressure to transfer the moment from acoustic wave to resident particles. This driving force transports the particles towards the center of a flow channel.[24] Evaporation-induced assembly depends on the rate of sedimentation of the particles. This method cannot be applied for particles which sediment at a slower rate than liquid evaporation. Furthermore, this process takes a long time (minutes to hours) to manipulate the particles.

Recently, a new technique, known as rapid electrokinetic patterning (REP) have emerged as an alternative method for concentration, patterning and sorting of colloidal particles.[25–28] REP utilizes a highly focused laser beam on the parallel-plate electrodes to generate localized hot-spot. An alternating electric field is applied between the electrodes with appropriate AC frequency. The hot-spot causes local gradient in electrical properties (conductivity and permittivity) which generate an electrokinetic force, commonly known as electrothermal forces. Toroidal vortex motion caused by electrothermal forces brings the particles at the illuminated spot. Using this technique nanoparticles are also manipulated on the electrode surface.[29] The method of opto-electrokinetic manipulation was adopted in many applications including biology and biomedical processes. Concentration, patterning, trapping, translation of microorganisms of bioassay system were efficiently performed using this technique.[30] Furthermore, based on the size of the microorganisms separation of a particular entity is possible under same landscape. Using REP it is also possible to enhance the sensitivity of detection of a bead-based bioassay system.[25,26] Combined effects of electrothermal flow with insulator-based dielectrophoresis can also trap micrometer and submicrometer particles efficiently.[31,32] For such a scenario, insulating structures generates Joule heating leading to electrothermal flow, which entrains colloidal particles to enrich at the vicinity of insulating tips.

Electrokinetic mechanisms often involve Joule heating, which is generated due to flow of electric current through the electrolyte.[33–36] Joule heating causes temperature gradient into the medium and affects the functionality of electrokinetic mechanisms and associated processes including biomedical, biochemical reactions.[37–40] Beyond a threshold temperature, properties of biological and biochemical samples are changed, and detrimental effects of Joule heating are observed.[41–44] To address the Joule heating effects and its consequences studies were conducted on electrokinetic flow with Joule heating.[45–49] Investigations were also carried out with different thermal conditions at the channel entrance.[50–52] Above studies on heat transfer characterizations are of conventional electroosmotic driven flow, where researchers reported their findings to minimize the adverse effects of Joule heating. Importantly, electrothermal flow is generated primarily due to Joule heating internally[53–55] or externally.[56,57] Therefore, Its operation uses the disadvantages of electroosmosis process. It has potential to transport fluid,[58,59] mixing of fluids,[60,61] and controlling the contact line dynamics of a binary system [62–64] efficiently and effectively.

The present work reports an alternative method of electrothermal flow, to manipulate and concentrate colloidal particles, which use internally evoked Joule heating to generate sharp temperature gradients, and thus, may be named as Joule heating-induced REP. To generate the temperature gradient towards a hotspot, optically-induced REP uses expensive infrared laser. Only 15% of the total power of the laser reaches the illuminated spot,[29] thereby wasting excessive thermal energy. Here, a similar hotspot is generated with a slight tweak in the chip-design. The bottom electrode is wrapped by an insulating layer and a very narrow hole is drilled in the insulating layer. Electrolyte gets in contact through the narrow drilled portion and concentrated Joule is generated at that location. Thus, a temperature gradient is generated in the fluid domain where the highest temperature is induced at the drilled area over the bottom electrode. The present technique is highly energy efficient, more integrable in LOC devices and does not deal with any externally employed component which provides thermal energy to the system. Further, laser-induced REP depends on the square of applied voltage whereas Joule heating-induced REP follows the trend of fourth power dependence on voltage. Thus, the present methodology of particle manipulation advocates an energy efficient technique compared to existing laser-induced REP.

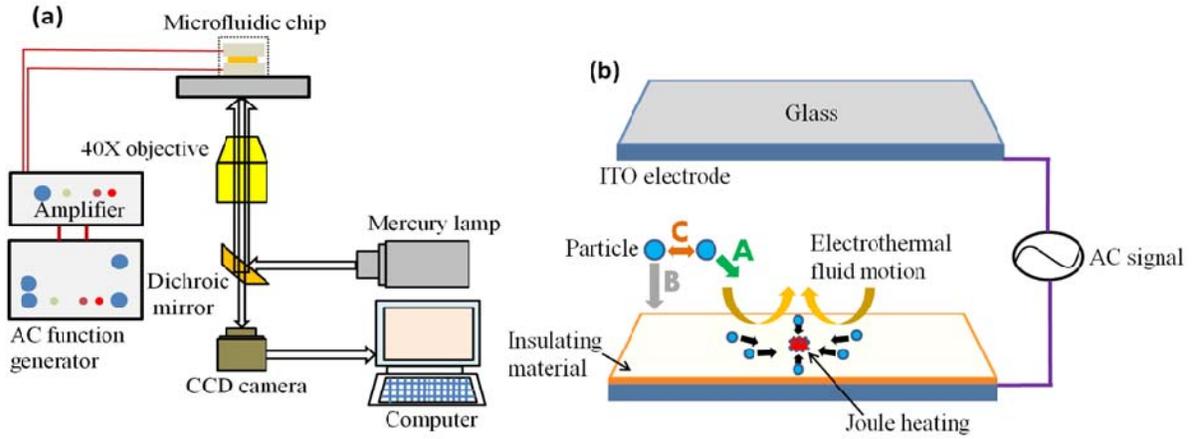

FIG. 1. (a) Schematic representation of experimental set up. (b) Illustration of Joule heating enabled Rapid electrokinetic patterning (enlarge view of microfluidic chip). Fluid with suspended particles is injected into the chip. AC electric field is imposed between the parallel electrodes. Joule heating is generated at the centre of the system where a hole is drilled in the insulating material. Electrothermal vortex brings particles at the hotspot.

**EXPERIMENTAL STUDIES**

**Materials**

The schematic of the complete set up is shown in the Fig. 1 ((a), (b)) that depict the microfluidic chip for Joule heating enabled REP mechanism. Two ITO-coated glass substrates separated with a double-sided adhesive tape (thickness: $35\,\mu m$) are connected with an AC power source. The signal from the AC function generator (33250A, Agilent) is amplified 50 fold by an amplifier (9200, High Voltage Wide Band Amplifier). In our experiment, we are restricted in the range of voltage (RMS value), $\varphi_{rms}$ < 9V and frequency > 100 KHz to avoid bubble generation. The bubble formation due to electrolytic reaction strongly depends on the applied voltage and frequency of the AC signal.[65] The applied frequency of the AC source should be higher than the inverse electrode reaction time and voltage must be below critical voltage (at which microscopic bubble formation occurs) to perform bubble-free particle manipulation.[65] Most of the results are shown at a voltage of 7 $V_{rms}$ chosen arbitrarily below $9\,V_{rms}$ to avoid bubble formation. Besides, we have not imposed frequency above 500 kHz owing to the limitation of availability of AC frequency function generator coupled with an amplifier.

Two parallel electrodes containing electrolyte generate uniform Joule heating which cannot induce thermal gradient in the fluid. To generate non-uniform temperature distribution and a concentrated heated region, the bottom electrode (thickness: 150nm) is wrapped using a thin insulating material (cello tape, thickness: 40 μm) and a hole (diameter = 50 μm) is drilled at the middle of the insulator. Due to the contact of the electrolyte at that particular drilled region, a hotspot is generated and temperature gradient is induced, which eventually forms the backbone of the design, as will be described below.

In our investigations polystyrene particles (Thermo Fisher Scientific) of four different sizes (diameter: 0.5, 1, 2, and 4 μm) are used. KCl (Merck Life Science Pvt. Ltd.) electrolyte solutions (conductivity 0.038 S/m) with suspended particles are introduced between the parallel plate electrodes. Particles are fluorescent red and are neutrally buoyant in the medium. The surface-charged, polystyrene particles act as polarizable and charged species, and electrothermally modulated particle aggregation characteristics of these microspheres are observed in this work by a fluorescence imaging technique. An inverted microscope (IX71, Olympus) is used to observe the particle aggregation. 40X and 10X objective lens, and a CCD camera (ProgRes MF$^{cool}$) are used to capture the images.

**Electrothermal effects**

The bottom electrode gets in contact with top electrode through the electrolyte placed on drilled portion of the insulating material. A highly non-uniform electric field is generated on activation of the AC signal. The non-uniformity depends on the diameter of the hole. A larger diameter passes the electric field across a large volume of the fluid and a uniform electric field will be imposed. However, electrothermal motion is governed by a sharp temperature gradient which is caused by a non-uniform electric field. We adopt a very small hole into the insulating layer to generate a non-uniform electric field and a sharp temperature gradient imposed from the non-uniform electric field. The strong thermal field introduces local variation in dielectric properties of the solution. To maintain the charge neutrality, free charges are induced into the bulk liquid, which causes electrothermal vortex.[66] The time-averaged electrothermal body force that drives the fluid is expressed by[67–69]

$$\mathbf{F_E} = -\frac{1}{2}\left[\left(\nabla\sigma/\sigma - \nabla\varepsilon/\varepsilon\right)\cdot\mathbf{E}\frac{\varepsilon\mathbf{E}}{1+(\omega\tau)^2} + \frac{1}{2}|\mathbf{E}|^2\nabla\varepsilon\right], \qquad (1)$$

where $\nabla\sigma/\sigma = (1/\sigma)(\partial\sigma/\partial T)\nabla T$ and $\nabla\varepsilon/\varepsilon = (1/\varepsilon)(\partial\varepsilon/\partial T)\nabla T$ are the conductivity gradient and permittivity gradient, respectively. $\mathbf{E}$ is the electric field. $\omega$ is the angular frequency of the AC signal. $\tau$ is the charge relaxation time of the fluid. The surrounding fluid moves towards the hotspot located on the drilled-hole of the insulator. The suspended colloids are transported to this trapped region (Fig. 1(b)-A). Other electrokinetic forces, such as dielectrophoretic force and electrostatic interactions interplay with van der Waals attraction that holds this aggregated colloidal particles near the electrode surface (Fig. 1(b)-B, 1b-C).[70,71] Dielectrophoresis forces are originated from the non-uniform electric field. In addition, interactions of the accumulated particles with the electric field generate non-uniformity in the electric field which also forms a source of dielectrophoresis forces.

**NUMERICAL STUDIES**

Alternating current electrothermal fluid flow is the coupled effect of electrical, thermal and velocity fields. Application of electric field induces temperature gradient in the media. Temperature gradients and electric field, combined together, impose bulk body force on the fluid. Under quasi-electrostatic condition (i.e., negligible magnetic field effect [72,73]), the imposed electrostatic field $\mathbf{E} = -\nabla \varphi$ is governed by [74,75]

$$\nabla \cdot (\sigma \nabla \varphi) = 0, \qquad (2)$$

where $\varphi$ is the voltage. For small temperature changes, variation in dielectric properties can be expressed as[67]

$$\varepsilon(T) = \varepsilon_0(T_0)(1 + \alpha T),$$
$$\sigma(T) = \sigma_0(T_0)(1 + \beta T), \qquad (3)$$

where $T$ is the local temperature and $T_0$ denotes the reference temperature. For aqueous solution, the gradients of dielectric properties are $\alpha = -0.004\,\text{K}^{-1}$ and $\beta = 0.02\,\text{K}^{-1}$.[61,76]

The non-uniform electric field generates thermal field through the induction of spatially-varying Joule heat into the medium. The energy equation which governs the temperature variation reads

$$\rho C_p \frac{\mathrm{D}T}{\mathrm{D}t} = \nabla \cdot (k \nabla T) + \sigma |\mathbf{E}|^2, \qquad (4)$$

where $\rho$ is the mass density of the fluid. $C_p$ is the specific heat and $k$ is the thermal conductivity. The Joule heat $\sigma |\mathbf{E}|^2$, which plays the key role in transporting the particle, is concentrated sharply above the drilled hole of insulating layer.

The induced flow, by virtue of its motion at the reduced scales, is laminar and incompressible. The continuity equation and the Navier-Stokes equation for fluid flow are written as

$$\nabla \cdot \mathbf{V} = 0, \tag{5}$$

$$\rho \frac{D\mathbf{V}}{Dt} = -\nabla p + \nabla \cdot \left[\mu\left(\nabla \mathbf{V} + \nabla \mathbf{V}^T\right)\right] + \mathbf{F_E}, \tag{6}$$

where $\mu$ is the fluid viscosity. $p$ is the pressure. $\mathbf{V}$ is the fluid velocity. The vortex motion to carry the colloidal particles towards the trapped region is controlled by the electrothermal forces $(\mathbf{F_E})$.

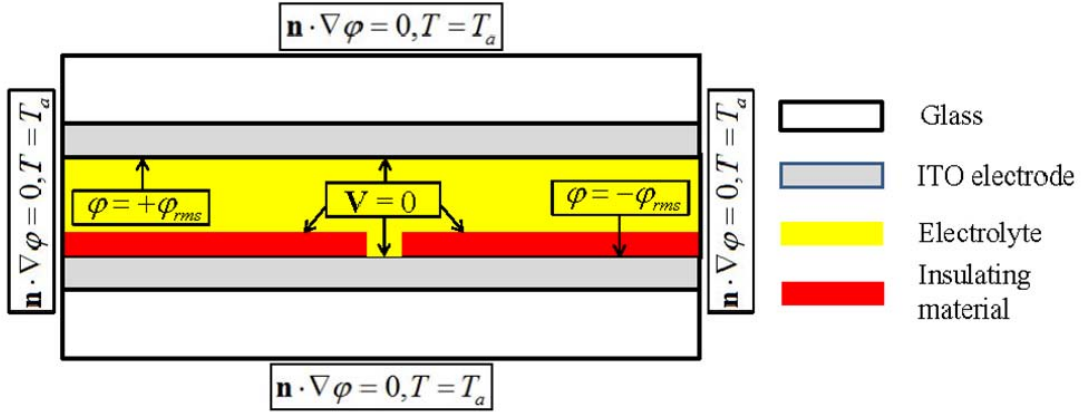

FIG. 2. Domain (front view) for numerical simulations. Boundary conditions are shown in the respective locations. The length and width of domain are $500\,\mu m$ and $500\,\mu m$, respectively.

The schematic of the physical system and boundary conditions are shown in the Fig. 2. The fluid domain occupies the space between the parallel electrodes. The suspended particles are injected into the fluid domain. This space along with the hole in the insulating layer is the flow domain whereas electrostatic equation and energy equation are solved in the entire setup. The physical properties of the fluid media are: $\rho = 1000\,\text{kg/m}^3$, $C_p = 4,184\,\text{J/kgK}$, $\mu = 0.00108\,\text{Pa s}$, $k = 0.6\,\text{W/mK}$, $\sigma = 0.038\,\text{S/m}$, and $\varepsilon = 7.08 \times 10^{-10}\,\text{C/Vm}$. Dielectric constant and thermal conductivity of glass and insulating material are: $\varepsilon_{r,\,glass} = 3.2$, $\varepsilon_{r,\,insulator} = 2.2$, $k_{glass} = 1.4\,\text{W/mK}$, and $k_{insulator} = 0.2\,\text{W/mK}$. The thickness of the ITO is negligible compared to other dimensions and hence in the simulation domain we do not consider the ITO. Electric field is imposed setting the voltage $\pm \varphi_{rms}$ on the electrodes whereas outer boundaries are insulated. At each interface, continuity conditions on the current are applied. To solve the energy equation outer boundaries are set at ambient temperature $T_0 = 298\,\text{K}$. We consider continuity of heat flux at fluid-insulator, insulator-electrode, and fluid-electrode interfaces. For velocity field, standard no slip and no-penetration boundary conditions are imposed on the electrode and insulator surfaces.

Three-dimensional simulations are performed using the commercial software package COMSOL Multiphysics to obtain the temperature distribution and velocity field. We consider only the electrothermal forces to play a leading role in colloidal clustering as it is the dominating force over other electrokinetic forces. The effect of buoyancy force can be neglected compared to other forces at reduced volume (characteristics length of the chamber $<100\,\mu m$).[29,77] Furthermore, colloidal aggregation in REP is not the effects of thermophoresis; the sole fluid motion is controlled by the electrothermal mechanism.[78] Mesh-independent solution is ensured by performing several simulations for different grid resolutions.

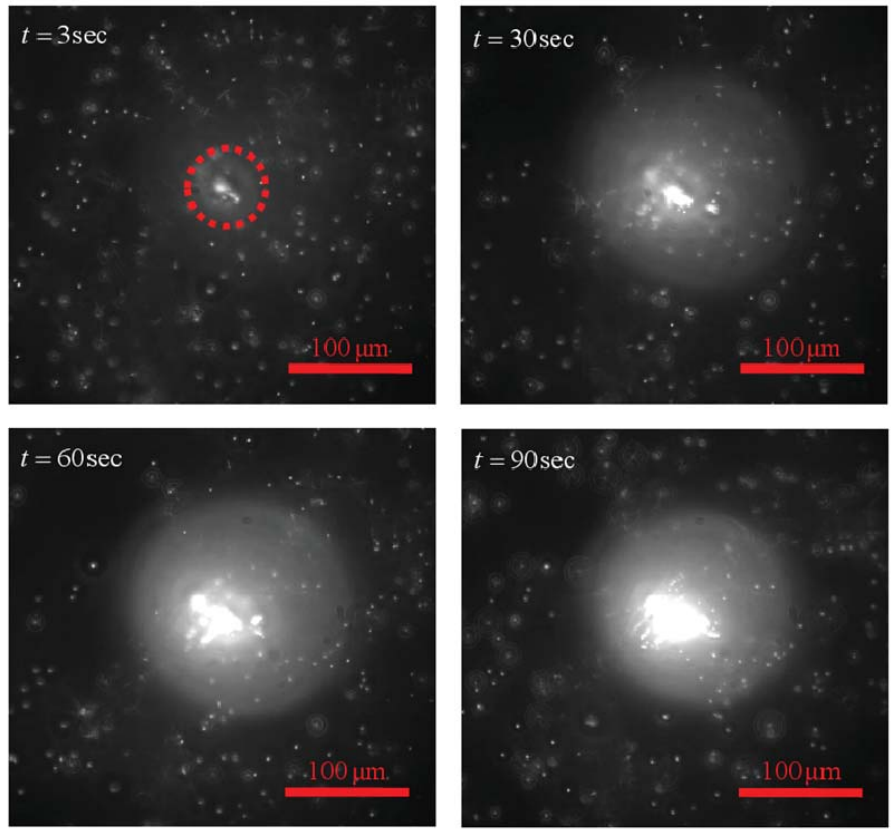

FIG. 3. Experimentally obtained particle (diameter: 1 μm) aggregation at different time instances ($t = 3\,\text{sec}$, $t = 30\,\text{sec}$, $t = 60\,\text{sec}$, and $t = 90\,\text{sec}$) for an AC voltage of $7\,\text{V}_{\text{rms}}$, a frequency of 500 kHz, and a electrical conductivity of 0.038 S/m.

**RESULTS AND DISCUSSIONS**

Fig. 3 shows the experimentally obtained time sequences of particle aggregation on the electrode surface for applied voltage $7\,V_{rms}$, frequency 500 kHz, and electrical conductivity 0.038 S/m. As the electric field is activated, the fluid body on the drilled portion (shown by dotted line in the first image) experiences strong electric field leading to intense Joule heating. The region of the highest temperature occurs at the centre of the hole in the insulating material. Temperature drastically reduces from the centre of the Joule heating area as one goes away from the hole. The sharp temperature gradient induces free charges by generating inhomogeneities in the electrical properties. The mobile charges bring the particles via toroidal vortex motion at the centre of the heating region. The aggregated particles are prone to leave the hotspot due to microfluidic vortex which generates fluid motion towards the outer region from the hot spot. However, particle-electrode holding electrokinetic forces traps the particles and enables particle concentration. As the time passes, the number of trapped particles increases and the area over which particles aggregate also becomes large. The layer by layer particle accumulation forms a stack. It is observed that light intensity around the insulating hole sharply increases for first 60s and gradually increases for rest of the time. The rate of particle aggregation decreases with time and cluster of particles are formed over the hotspot. A movie (Movie1) showing the particle aggregation process is provided in the Supplementary material for further understanding of the Joule heating-induced REP mechanism.

Point to be noted here is that to generate directional flow (surrounding the hotspot) for particle aggregation, size of the hole in the insulating material should not be large. Such a configuration may generate some electrothermal flow without particle aggregation. Movie2 in the supplementary information highlights such a scenario where a large hole ($600\,\mu m$) above the bottom electrode does not allow temperature gradient along a radial direction, so that vortex motion cannot bring the surrounding fluid towards the central point. Due to arbitrary thermal field, the fluid flow occurs in the large hole-region above the bottom electrode without showing any particle aggregation.

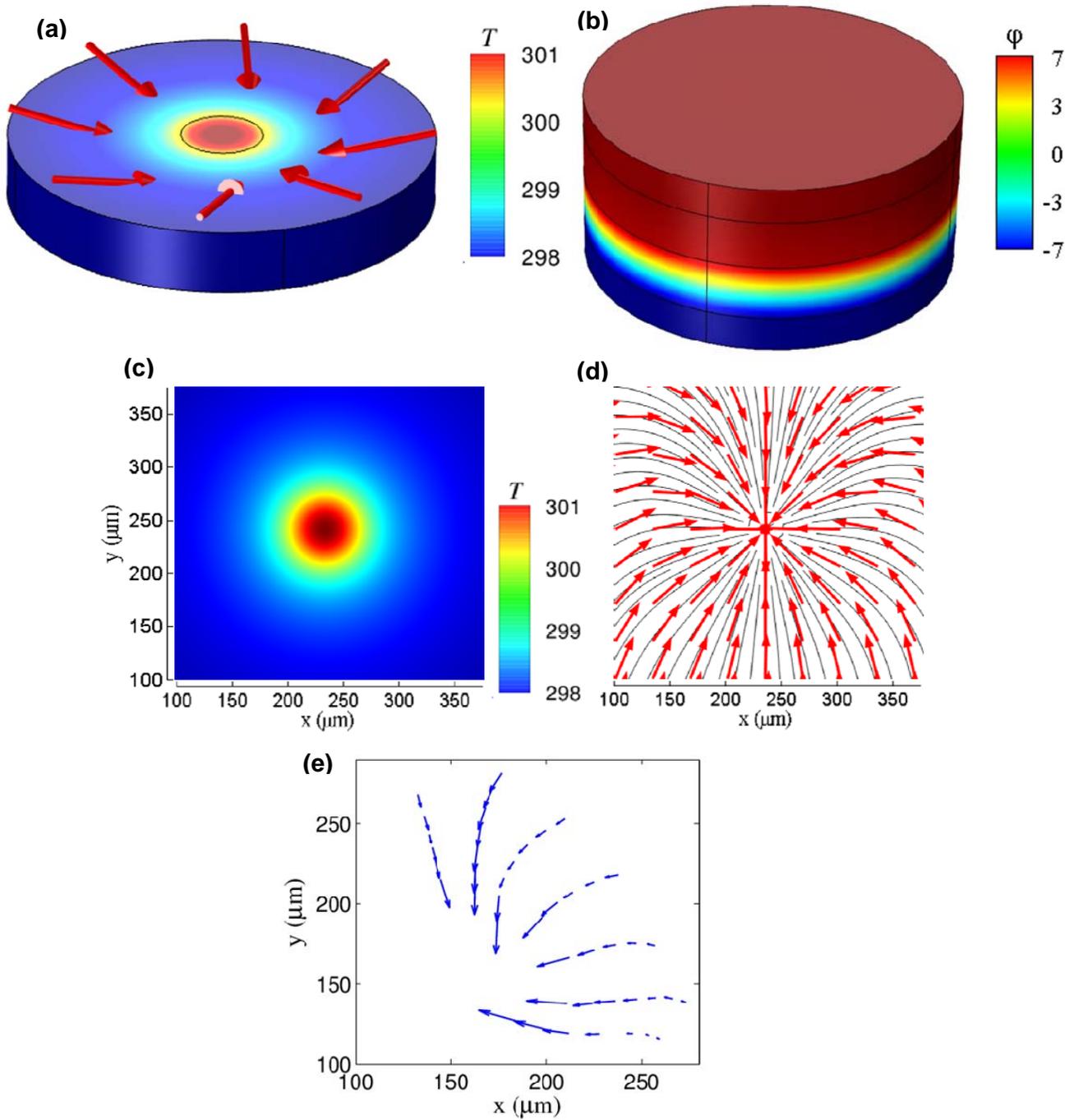

FIG. 4. (a) Depicts 3D distribution of temperature and arrow lines of the velocity vector. (b) Shows 3D distribution of potential. The 3D disk height for (a) and (b) are 50 μm and 125 μm, respectively, and are taken above the bottom electrode and from 25 μm below the bottom electrode. Disk diameter is 350 μm. (c) Simulated results of surface plot of temperature and (d) streamline contours at 10 μm above the insulating material. The surface plots are the top surface of 3D disk of (a). (e) Experimentally obtained Vector plot of fluid motion. We selectively track seven particles to understand the flow field. This vector plot is similar to upper right quadrant of the numerically obtained vector plot (d) and matches qualitatively with numerical results. For all plots a voltage of $7\,\text{V}_{\text{rms}}$, a frequency of 500 kHz, and electrical conductivity of 0.038 S/m are adopted.

The flow physics associated with Joule heating enabled particle concentration can be further understood by the temperature and streamline distributions obtained from our numerical results (Fig. 4, 2D and 3D). The parameters used in the simulations and details of the plots are mentioned in the figure caption. It is clear that a concentrated Joule heat is generated at the centre of the domain where fluid has contact with electrodes. Strong electric field generates Joule heat at the fluid centre. Due to decrease in electric field strength as one moves away from the central hole location, the temperature also decreases. It is noteworthy that temperature gradient in the domain is very sharp owing to a sudden drop in temperature from this central location. Strong electrothermal forces, generated by the conductivity and permittivity gradient caused by the sharp gradient in temperature, bring the surrounding fluid to the central location. Streamline plot in figure 4(b) shows the flow patterns. It is clear that bulk fluid flows towards the central location and carries the particles to the hotspot. It is obvious that to maintain the continuity of mass, fluid from the centre at other plane moves away from the hotspot. However, particle-electrode holding force does not allow particles to move with the outward vortex. Hence, particle aggregation takes place at the centre where insulating layer is drilled. The motion of the fluid towards the centre hole exhibits similar motion with experimentally obtained fluid flow. We have shown the tracking details of some particles in Fig. 4(e), which shows a clear scenario of convergence of the fluid motion towards a centre location obtained numerically at right quadrant of the Fig. 4(d). We also show the potential distribution in Fig. 4(b) where a height of 25 $\mu m$ of total height ($125\,\mu m$) of the disk is in the bottom electrode and also in top electrdoe. From the figure, it is clear that the positive potential of the top electrode is distributed in most of the part of the disk although the portion of the top electrode and bottom electrode is same in the disk. This is because aqueous solution (height $35\,\mu m$) of higher conductivity is located adjacent to upper electrode and insulating material (height $40\,\mu m$) of lower conductivity is located adjacent to the bottom wall. Conducting solution assists to distribute the potential uniformly in the media. On the other hand, insulator resists in potential distribution above the bottom electrode.

Particle accumulation process is further characterized by the intensity variation with time. Fig. 5(a) shows the maximum fluorescence intensity as a function of time. The fluorescence intensity is normalized with respect to the maximum intensity of 192

(arbitrary unit) occurred at time 90s. To measure the fluorescence intensity at different time instances, reference intensity (intensity at $t = 0$) is subtracted from intensity of all time instances. The light intensity sharply increases for first 60 seconds and gradually increases for the rest of the time and almost saturated at the end of the accumulation. With time, number of accumulated particles increases and fluorescence intensity increases. However, due to saturation of aggregation process with time, fluorescence intensity also attains a plateau.

The effects of voltage on the rate of particle aggregation are shown in the Fig. 5(c) and 5(d), where Fig. 5(c) depicts the average velocity at the periphery of the light ring (shown by dotted line in Fig. 5(b)) and Fig. 5(d) shows the data of local velocity as a function of the radial position. Centre of the particle clustering is taken as the origin.

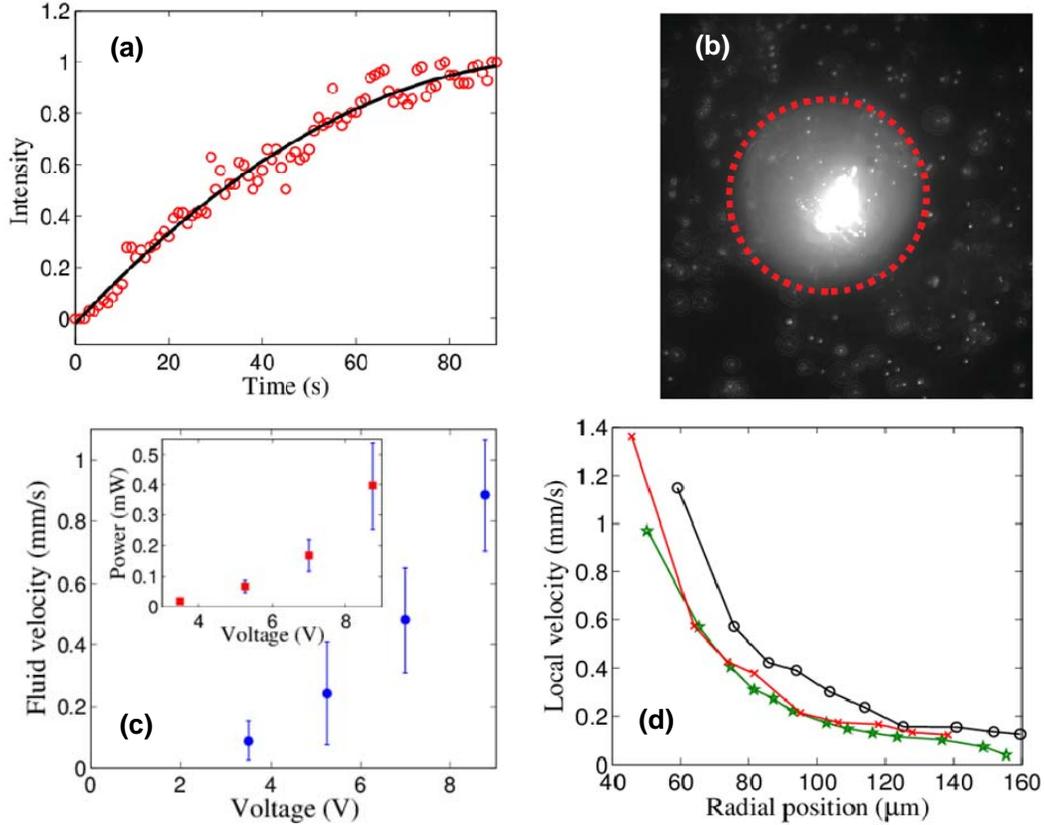

FIG. 5. (a) Rate of particle concentration in the form of fluorescence intensity. AC voltage for these set of data is $7\,V_{rms}$ (b) Image depicts locations where average velocity was measured. (c) Variation of average velocity as a function of voltage at the periphery of the light ring. Inset plot shows power budget to execute the ACET mechanism. (d) Variation of local velocity along the radial direction at three different locations for voltage $7\,V_{rms}$. Centre of the clustered particles is taken as origin (i.e., radial coordinate $r = 0$). All plots are experimentally obtained for the imposed frequency and electrical conductivity of 500 kHz and 0.038 S/m, respectively, and particles of diameter 1 μm are taken.

The applied voltage for variation of local velocity is taken as $7\,V_{rms}$. The mean velocity of the particle rapidly increases with the applied voltage. This is consistent with reported electrothermal-based micropumps.[58,59,79] Higher the fluid velocity, faster will be the process of particle aggregation. The electric field strength at the central location increases with voltage and stronger non-uniform electric field is generated. This, in turn, enhances the electrothermal body forces. Toroidal vortex motion becomes stronger at increased electrothermal forces. Ideally, fluid velocity in an electrothermal process varies[80] as $u \sim \phi^4$. Through a regression analysis of velocity data, an index 2.6 of velocity dependence on voltage is obtained. This is slightly lower than the theoretically obtained value. The reason for such deviation may be attributed with the fact that other electrokinetic forces, such as dielectrophoresis forces, electroosmosis forces may interplay with electrothermal forces at different scales. However, impacts of these forces are not significant compared to electrothermal forces.[81] Fluid velocity in a laser-induced REP is proportional to the square of the AC voltage.[29] Interestingly, Joule heating-induced REP velocity follows the fourth power dependency on the applied voltage. In the body force term, one can see flow velocity scales with the square of the electric field. Accordingly, laser-induced REP varies with the square of voltage. However, in a Joule heating enabled REP system, electric field generates the Joule heat. A balance between diffusion term and Joule heating term in the energy equation gives $\Delta T \sim E^2$. Since Joule heating-induced body force is $F_E \sim E^2 \nabla T$,[82] dependencies of fluid velocity on electric field strength and voltage are $u \sim E^4$ and $u \sim \phi^4$, respectively. This is a big advantage of Joule heating-induced particle aggregation over laser-induced REP. On one side, the present technique does not demand any external light source to generate the non-uniform thermal field; while on the other side, this method provides higher velocity to bring the particle rapidly to the desired location and thereby resulting in faster particle aggregation.

Power budget is estimated and shown in the inset of Fig.5(c) Power requirement is low (below 1 mW) to drive the system. laser-induced REP generally uses laser beam of power 20 mW.[30] In addition, it uses an electrical power, order of which is almost same as that used in the present study. Therefore, due to absence of external light source, the present arrangement is highly energy efficient. Local velocity of the particles slowly increases first and then sharply increases as the distance becomes closer to the centre of the particle clustering (Fig. 5(d)). Electric field strength and temperature gradient both increase at a higher rate close to hotspot and electrothermal forces become stronger at that location. As a result, particles move with higher velocity.

Electrode spacing also affects the particle concentration process. Channel between the electrodes was made by fixing double-sided adhesive tape at the edge of the insulating layer. So far, all the experiments were performed fixing a single tape between the electrodes, which builds a electrode gap of $75\,\mu m, (=35+40, 40\,\mu m$ for inslulating layer). Experiments were also performed changing the Electrode gap i.e., height of the channel placing more number of tapes between the electrodes. For one tape, the velocity of the fluid shows a magnitude of 0.88 mm/s (applied voltage 8.75 V). However, for two tapes (thickness of 110 μm) velocity drops to 0.17 mm/s. As mentioned before that power law dependence of fluid velocity on electric field strength is $u \sim E^4$. On the other hand, relation between voltage, gap and electric field is $E \sim \varphi/g$, $g$ is the electrode gap. Therefore, for a given applied voltage electric field strength decreases with decreasing electrode spacing. Hence, fluid velocity also decreases with decreased electrode spacing. The particles are concentrated at lower rate for $g = 110\,\mu m$. When we added one more tape between the electrode ($g = 145\,\mu m$) the rate of particle aggregation almost becomes negligible.

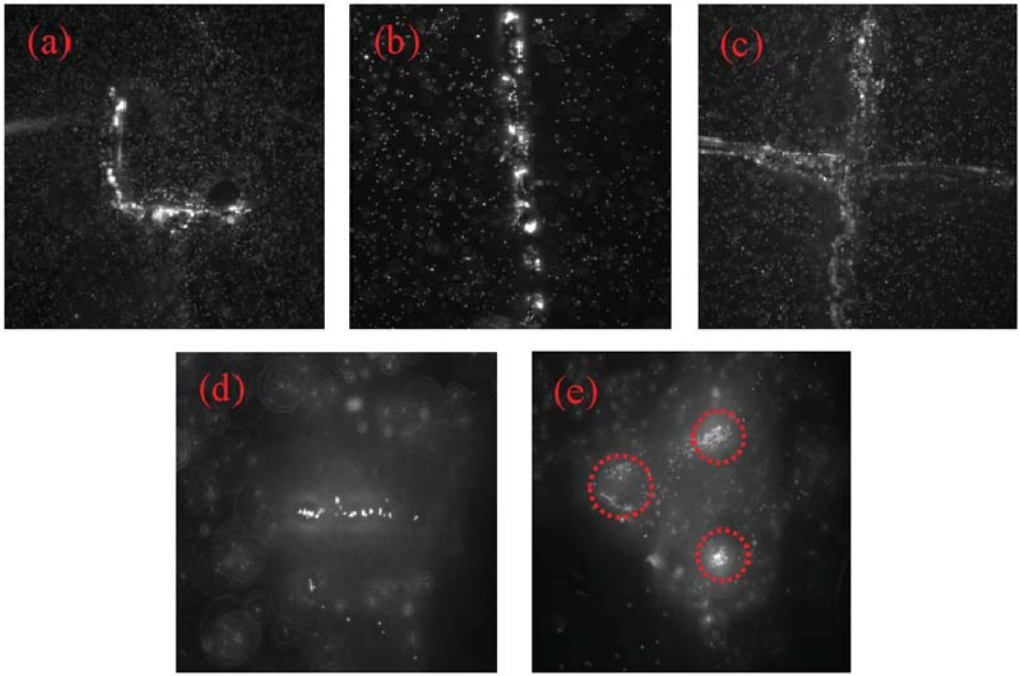

FIG. 6. Particles of diameter 1 μm are trapped at desired locations to form different clustering structures: (a) "L", (b) "I", and (c) "+" shapes. (d) Stacking of particles along a line can be formed. (e) Controlling the Joule heating at different spots particles can be aggregated at different locations (shown by dotted line). For all subfigures applied voltage, frequency and electrical conductivity are $7\,V_{rms}$, 500 kHz and 0.038 S/m, respectively. All the images are taken from experiments. The structures of the patterned are same as structures of the holes. The widths of the holes are in the range of 50-200 μm

Particle aggregation process further depends on the shape of the hole on the bottom electrode. Accordingly, Joule heat pattern will be generated resembling the shape of the opened portion on the bottom electrode. Fig. 6 shows the different patterns of particle groups: "L", "I", "+" shapes which resemble the pattern formed on the insulating layer. On the other hand, particles can be stacked in a line (Fig.6(d)). In addition, by generating multiple hotspots, particles can be accumulated over the electrode surface at different places (Fig. 6(e)). All the images are captured at voltage $7\,V_{rms}$. Thus, the particulate accumulation follows the Joule heat pattern, since it tends to accumulate at the section of peak heating regions, which, in turn, follows the drilling or opening shape on the bottom electrode. The manipulation of the fluid dynamics and particle aggregation regions through simple designing of the insulating layer testifies the elegance of the present method.

## CONCLUSIONS

We have experimentally demonstrated AC electrokinetic technique based on generating non-uniform Joule heating for particle aggregation and pattern formation of particle groups. The coupling between electric and thermal fields generates toroidal vortex motion where surrounding fluid approaches a targeted hotspot location and particles are carried into that spot to form a particle cluster. Our designed microfluidic chip for particle manipulation process is highly energy saving compared to laser-induced particle manipulation. Further, patterning the heat generation distribution, different structures of accumulated particles can be formed.

## SUPPLEMENTARY MATERIAL

See supplementary material for Movies of particle accumulation (Movie1) and no particle accumulation (Movie2) scenarios at voltage $7\,V_{rms}$.

## ACKNOWLEDGEMENTS

This research has been supported by Indian Institute of Technology Kharagpur, India [Sanction Letter no.: IIT/SRIC/ATDC/CEM/2013-14/118, dated 19.12.2013].